\documentclass[useAMS,usenatbib,usegraphicx]{mn2e}
\usepackage{amsmath}
\usepackage{amssymb}
\usepackage{xspace}

\newcommand{\wmap}{\emph{WMAP}\xspace}

\def\ltsim{\ifmmode\stackrel{<}{_{\sim}}\else$\stackrel{<}{_{\sim}}$\fi}
\def\gtsim{\ifmmode\stackrel{>}{_{\sim}}\else$\stackrel{>}{_{\sim}}$\fi}

\newcommand{\beq}{\begin{equation}}
\newcommand{\eeq}{\end{equation}}

\title[Synchrotron, Cosmic Rays, and Magnetic Fields]{Connecting Synchrotron, Cosmic Rays, and Magnetic Fields in the Plane of the Galaxy}
\author[T. R. Jaffe et al.]{T. R. Jaffe $^{1,2}$\thanks{E-mail:tess.jaffe@cesr.fr}, 
   A. J. Banday$^{1,2,3}$\thanks{E-mail:Anthony.Banday@cesr.fr}, 
   J. P. Leahy$^{4}$\thanks{E-mail:j.p.leahy@manchester.ac.uk},
   S. Leach$^{5,6}$\thanks{E-mail:leach@sissa.it},
  A. W. Strong$^{7}$\thanks{E-mail:aws@mpe.mpg.de}\\
$^{1}$ Universit\'e de Toulouse; UPS-OMP; IRAP;  Toulouse, France\\
$^{2}$ CNRS; IRAP; 9 Av. colonel Roche, BP 44346, F-31028 Toulouse cedex 4, France\\
$^{3}$Max Planck Institute for Astrophysics, Karl-Schwarzschild Str. 1, D-85741 Garching, Germany\\
$^{4}$Jodrell Bank Centre for Astrophysics, School of Physics and Astronomy, The University of Manchester, Oxford Road, \\
\hspace{1cm} Manchester, M13 9PL, United Kingdom\\
$^5$SISSA, Astrophysics Sector, via Beirut 2-4, I-34014 Trieste, Italy.\\
$^6$INFN, Sezione di Trieste, I-34014 Trieste, Italy.\\
$^{7}$Max-Planck-Institut f\"ur Extraterrestrische Physik, Postfach 1312, D-85741 Garching, Germany \\
}
\begin{document}

\date{}

\pagerange{\pageref{firstpage}--\pageref{lastpage}} \pubyear{}

\maketitle

\label{firstpage}

\begin{abstract}
  We extend previous work modeling the Galactic magnetic field in the
  plane using synchrotron emission in total and polarised intensity.
  In this work, we include a more realistic treatment of the
  cosmic-ray electrons using the {\sc galprop} propagation code optimized to
  match the existing high-energy data.  This addition reduces the
  degeneracies in our previous analysis and when combined with an
  additional observed synchrotron frequency allows us to study the
  low-energy end of the cosmic-ray electron spectrum in a way that has
  not previously been done.  For a pure diffusion propagation, we find
  a low-energy injection spectrum slightly harder than generally
  assumed; for $J(E)\propto E^\alpha$, we find $\alpha=-1.34\pm 0.12$,
  implying a very sharp break with the spectrum above a few GeV.  This
  then predicts a synchrotron brightness temperature spectral
  index, $\beta$, on the Galactic plane that is $-2.8<\beta<-2.74$
  below a few GHz and $-2.98<\beta<-2.91$ up to 23~GHz.  We find that
  models including cosmic-ray re-acceleration processes appear to be
  incompatible with the synchrotron data.
\end{abstract}


\section{Introduction}\label{intro}

Studies of both cosmic rays and magnetic fields in the Galaxy have
independently gained momentum recently from the advent of newly
available data.  The Fermi satellite has recently provided the most
precise direct measurements to date of the cosmic ray electron (CRE)
spectrum near the Earth \citep{fermi:2010} as well as of the
$\gamma$-ray sky that provides indirect measurements of the cosmic ray
distribution in the Milky Way \citep{abdo:2009b}.  Recently expanded
catalogs of Faraday rotation measures (RMs) by \citet{taylor:2009} and
\citet{vaneck:2011} (and soon GALFACTS\footnote{{\tt
    http://www.ucalgary.ca/ras/GALFACTS/}}) are allowing large-scale
magnetic field theories to be tested and rejected by the hugely
increasing amounts of data.  And last but not least, the first
full-sky maps of polarised synchrotron emission in microwave
frequencies provided by the Wilkinson
Microwave Anisotropy Probe (\wmap) (and soon the Planck
satellite\footnote{\tt http://www.esa.int/planck}) are beginning to
allow us to disentangle the various degeneracies that dog studies of
how the cosmic rays and the magnetic fields interact in the
interstellar medium (ISM).

In \citet{jaffe:2010} (hereafter Paper I), we studied the components
of the Galactic magnetic field using the three complementary datasets
of total synchrotron intensity, polarised synchrotron intensity, and
rotation measure.  In that paper, our aim was to determine the
relative strengths of the magnetic field components which we define as
coherent, ordered, and random.  These components can be separated by
examining Faraday rotation measure to fix the coherent component
and synchrotron emission to explore the ordered and random components.
The geometry allows us to separate them using the combination of
synchrotron polarised and total intensity.  See Fig.~1 from Paper I.

Ideally, we would have both such datasets at the same
frequency.  But we are attempting to determine the magnetic field
structure in the plane of the galaxy, which brings in two
complications.  At frequencies below a few GHz, we cannot use
polarised emission as a tracer of large-scale structure because of
Faraday depolarisation; the turbulent ISM effectively imposes a
polarisation horizon beyond which the polarisation signal is erased
\citep{uyaniker:2003}.  Above frequencies of a few GHz, however, the
total intensity is dominated on the Galactic plane by free-free
emission and the anomalous dust-correlated emission (generally
believed to be electric dipole emission from spinning dust grains).
It is not currently possible, therefore, to determine the magnetic
field components via comparison of polarised and total intensity at
the same frequency.  In Paper I, we used a frequency for total
intensity that is low enough that the synchrotron dominates, and
another frequency, well above the Faraday regime, for the polarised
intensity.

The drawback of this approach is the assumption of a simple power law
distribution of cosmic ray electrons, $J(E)\propto E^{-p}$, over all
energies with an index of $p=3$.  Given that the data span the range
from the low-frequency radio at 408~MHz to the microwave at 23~GHz,
this assumption is likely to introduce errors into the resulting
parameters inferred for the magnetic field strength.  Since two
synchrotron observables I and PI allow us to determine the relative
strengths of the three magnetic field components defined in Paper I,
with an additional low frequency total I dataset we can additionally
constrain the low-energy end of the cosmic ray electron 
spectrum.  The first aim of this paper is to see what can be
confidently measured using the available data.
  
It is often assumed that the low-energy CRE spectrum, i.e. below
$\sim4$~GeV, is harder than the spectrum in the Fermi frequencies,
where the CRE spectrum has been well measured between 7~GeV and 1~TeV
\citep{fermi:2010}.  Estimates of the low-energy spectral index have
been done with low-frequency radio surveys (e.g.,
\citealt{guzman:2011}), but lower than 408~MHz, one must begin to
worry about and model free-free absorption, which is particularly
difficult in the plane.  Direct local measurements near the earth of
CREs at energies below a few GeV are affected by solar modulation.
Therefore, it is the synchrotron emission below a very few GHz that is
likely the best method for measuring this spectrum, at least as an
average in the Galactic plane.  We test two possible datasets for the
intermediate frequency: the 1420~MHz full-sky survey of
\citet{reich:1982,reich:1986} and the 2326~MHz southern survey of
\citet{jonas:1998}.

Rather than assuming that the cosmic ray spectrum follows even a
simple broken power law, we use a complete cosmic ray propagation
code, {\sc galprop}\footnote{{\tt http://galprop.stanford.edu}}.  This
allows us to use physically motivated models of cosmic ray electron
and positron propagation -- including treatment of diffusion,
synchrotron energy losses, the local interstellar radiation field,
re-acceleration, secondary production, etc.; see \citet{strong:2007}
-- which are consistent with the Fermi data \citep{strong:2010}.
{\sc galprop} can give us a full four-dimensional model of the Galactic
cosmic ray distribution both as a function of Galactic position and of
energy.  We can then integrate the synchrotron emission not only over
the line-of-sight but also over the cosmic ray spectrum in order to
get the best possible prediction of the synchrotron emission.

As in Paper I, we use a full Markov Chain Monte Carlo (MCMC) analysis to
explore the parameter space of both the Galactic magnetic field and
now also the low energy cosmic ray electron injection spectrum.  We
have integrated the {\sc galprop} CRE propagation code into our {\sc
  hammurabi}\footnote{{\tt http://www.mpa-garching.mpg.de/hammurabi/}}
\citep{waelkens:2009} simulation code, now performing the full
integration over the electron spectrum rather than simply assuming an
electron index of $p=3$ as in Paper I.  We now obtain a result that is
fully self-consistent, in the sense that the cosmic rays are
propagated through a galaxy model that includes our magnetic field
model (which changes the CRE spectrum through synchrotron energy
losses) and then those CREs are used to determine the total observed
synchrotron emission.

The result is more reliable measurement not only a of the magnetic
field parameters but also of the low-energy cosmic ray spectrum than
is possible with other methods.  We can then compare our result to 
low-energy cosmic ray data from local measurements and comment on the
issue of solar modulation.  


\section{Observations}\label{sec:obs}

\begin{figure}
%
%
\includegraphics[width=\linewidth]{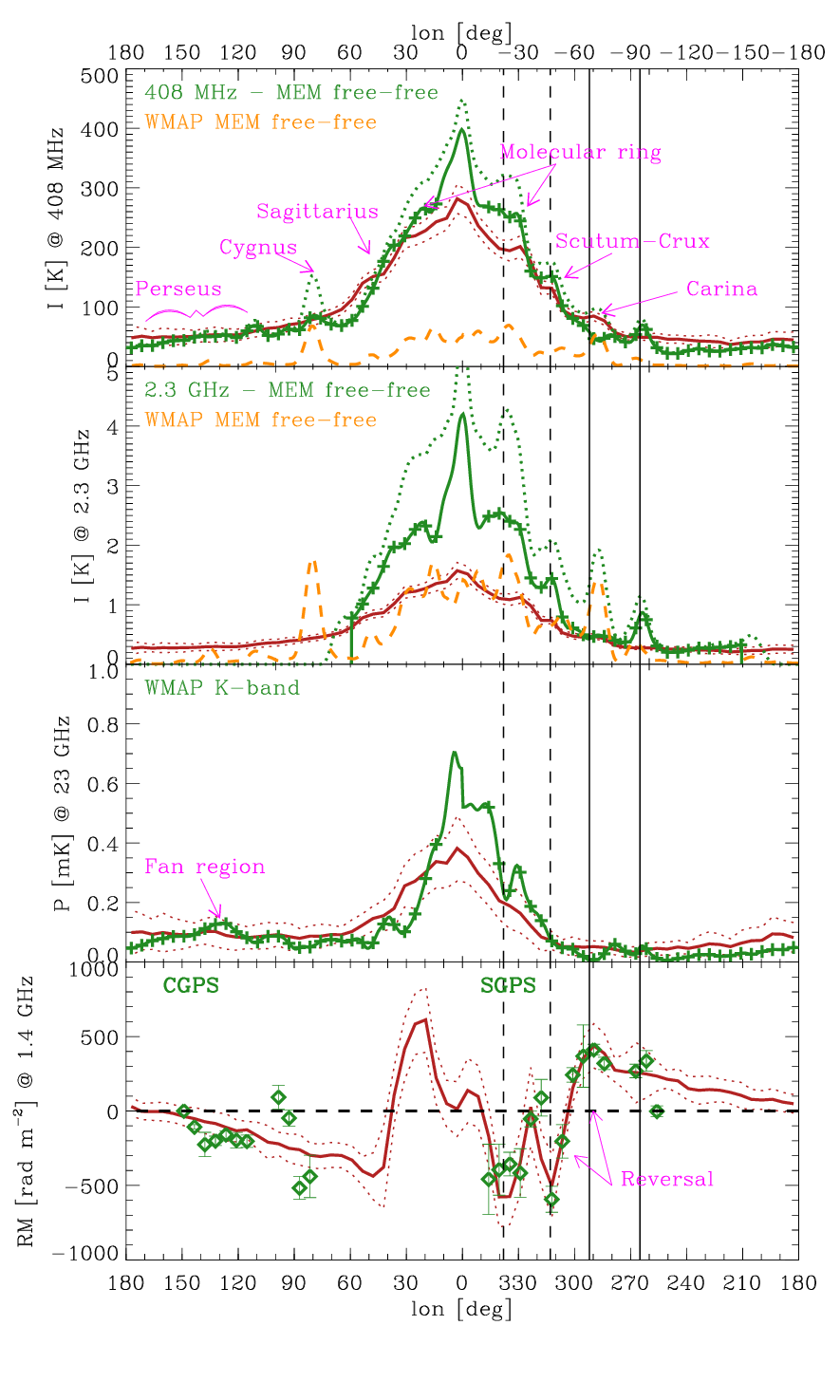}
\caption{The available observables ({\it green}) tracing the Galactic
  magnetic field.  The over-plotted model ({\it red}) is the original
  model from Paper I, while the data have been updated somewhat.  See
  text in \S~\ref{sec:obs}.  Furthermore, we have added the 2.3~GHz
  frequency which clearly shows that the power-law CRE spectrum does
  not match all of the data.  (The {\it dotted green} line is the raw
  data, while the {\it solid} is that after a free-free estimate, show
  in {\it dashed orange}, is subtracted.  See \S~\ref{sec:ffcor}.)}
\label{fig:plane_data}
\end{figure}

The data we use are shown as green solid lines in
Fig.~\ref{fig:plane_data} with annotations to point out interesting
features:
\begin{itemize}
\item {\em Top:} from \citet{haslam:1982}, the synchrotron
  total intensity at 408~MHz.
\item {\em Second:} from \citet{jonas:1998}, the synchrotron total
  intensity at 2.3~GHz.
\item {\em Third:} from \citet{Jarosik:2011}, the synchrotron
  polarised intensity at 23~GHz from the \wmap seven-year analysis.
\item {\em Bottom:}  the RM data from \citet{brown:2003,brown:2007}
  averaged into roughly $6\degr$ bins.  
\end{itemize} As described in Paper I, the sky maps are smoothed
  to a common low resolution FWHM of 3$\degr$.  The profile at zero
  latitude is then extracted and an additional boxcar smoothing
  applied to a resolution of roughly 6$\degr$.

With the exception of the second panel down, the datasets in
Fig.~\ref{fig:plane_data} are those used in Paper I with small
modifications.  For example, the \wmap data have been updated with the
seven-year results.  The free-free correction for the synchrotron
total intensity has also been modified, as described in
\S~\ref{sec:ffcor}.  The second profile from the top is an additional
synchrotron total intensity dataset at an intermediate frequency.  (It
is processed identically to the Haslam et al. data; see Paper I.)  It
is the addition of this dataset that will allow us to explore the
cosmic ray spectrum rather than to assume a simple power law over all
energies.  

The unsurprising fact that our model from Paper I (shown in red)
clearly does not fit this profile is the motivation for this work.
That model is based on a simple exponential disc for the spatial
distribution of CREs (which partly determines the longitude profile's
shape) and a single power law for the spectral distribution (which
determines the relative amounts of emission at each frequency).  The
failure of the model is largely due to the over-simplistic spectrum,
as we now have three frequencies with which we probe the cosmic ray
spectrum, and it is not expected to be a simple power law.  

Tangents to possible spiral arms are indicated, and vertical lines
mark interesting sight-lines (solid for positive RM, dashed for
negative), also shown on Paper I Fig.~4.  
The dashed orange line is
the estimate for the free-free contamination which has been subtracted
from the total emission at 408~MHz (dotted green line); see
\S~\ref{sec:ffcor}.

We note that other radio surveys on the plane exist (see, e.g. the
Bonn Survey Sampler\footnote{{\tt
    http://www.mpifr-bonn.mpg.de/survey.html}}), but the two radio
bands chosen are most appropriate for our purposes. A low frequency
gives a longer lever-arm for studying the CRE spectrum and also
minimizes the thermal emission on the plane.  At 408~MHz, the
synchrotron is sampled at sufficiently low a frequency but not so low
as to be affected by absorption effects on the plane.  At intermediate
frequencies, there are several surveys that include the galactic
plane, but many do not have sensitivity to large angular scales.  We
have also considered the 1420~MHz survey of \citet{reich:1982,reich:1986}.
 We find, however, that on the plane this survey is not consistent
  with the 2.3~GHz survey.  Though there are issues with both
  surveys, we have not found a clear explanation for this discrepancy.
  There is a potential issue with the 2.3~GHz survey in that it it
  sensitive to only one polarization direction.  This may over or
  under estimate the total intensity depending on the orientation of
  the polarization relative to the detector.  Given the significant
  depolarization at this frequency \citep[for example]{duncan:1997},
  we expect this effect to be quite small, of order a few percent at
  most.  Furthermore, we note that the 2.3~GHz survey is independently
  calibrated on a large-angular scale drift scan measurement at 2 GHz
  \citep{jonas:1998}.  We therefore consider the 2.3~GHz survey a
  better choice for this analysis on the galactic plane.

Since Paper I, new RM data have been published by \citet{vaneck:2011}.
These data cover part of the missing region of the plane most of
interest to attempts to model the coherent field structure.  The
purpose of this paper, however, is simply to modify our previous
analysis using more realistic CRE spectral models and an additional
synchrotron total intensity frequency.  To modify the coherent field
model to match the van Eck et al. data is therefore beyond the current
scope, though clearly we plan to incorporate the data in future work.
It is interesting to note that our proposed best-fitting spiral model
does not fit the new data, which we discuss more in the context of our
results in \S~\ref{sec:results}.  But this inconsistency is only
  in the positions of the arms, not in the strength of the magnetic
  field components.  It therefore does not have an impact on the
  current aim to constrain the CRE spectrum using the synchrotron
  spectrum.

\subsection{Offsets}\label{sec:offsets}

In Paper I, we treated all maps identically and fit an offset using a
$\csc(|b|)$ model to the sky outside of the plane and subtracting that
offset.  At 408~MHz, this gives a surprisingly high offset,  
which though not particularly significant for the analysis we did is
certainly much higher than the offset of 5.89~K \citep{lawson:1987} or the offset
found by \citet{reich:1988} of 2.7~K (based on comparison to their
1420 survey).

Clearly, the synchrotron sky is not well represented by a cosecant
law, as the matter distribution is not well approximated by a simple
slab model.  (Furthermore, depending on how it's done, it often leaves
unphysical negative pixels.)  Despite its shortcomings,
\citet{bennett:2003b} use this method as well and find that the Lawson
et al. offset of 5.89~K is sufficient and that their cosecant fits
show no evidence of an additional offset.  
Though there are a variety of
  other determinations of this survey's offset, 
for this work, we have decided to use the corrections given by
\citet{lawson:1987} for the 408~MHz survey, i.e. we subtract an offset
of 5.89~K.   Note that at this frequency, the offset uncertainty
  is only of order 1\% of the signal in the plane, and had we chosen
  the \cite{reich:1988} offset, our scientific results would be effectively
  unchanged.  We apply no additional offset correction to the 2.3~GHz
survey, as the temperature scale of the survey data has been set by
comparison to an absolutely calibrated survey at 2~GHz and had a
prediction for extragalactic contributions removed; see
\citet{jonas:1998} for details.


When also examining the free-free subtraction in the next section, we
fit 408~MHz as a template to the 2.3~GHz synchrotron profile with a
linear fit that essentially gives us a template scale factor based on
the structure and a measure of the implied offset between the
  surveys, similarly to a TT-plot.  As shown in
Fig.~\ref{fig:jonas_reich}, the fitted offset is very
small and well within the quoted uncertainties.

Note that the free-free model (see \ref{sec:ffcor}) has also been
examined for an offset using cosecant fits.  That analysis gives
offsets too small to be of significance: of order 0.07 mK at 33~GHz,
which is is too small to be seen in in a plot such as
Fig.~\ref{fig:plane_data} and is much smaller than any of the other
uncertainties.

\subsection{Processing}\label{sec:proc}

\begin{figure}
\includegraphics[width=\linewidth]{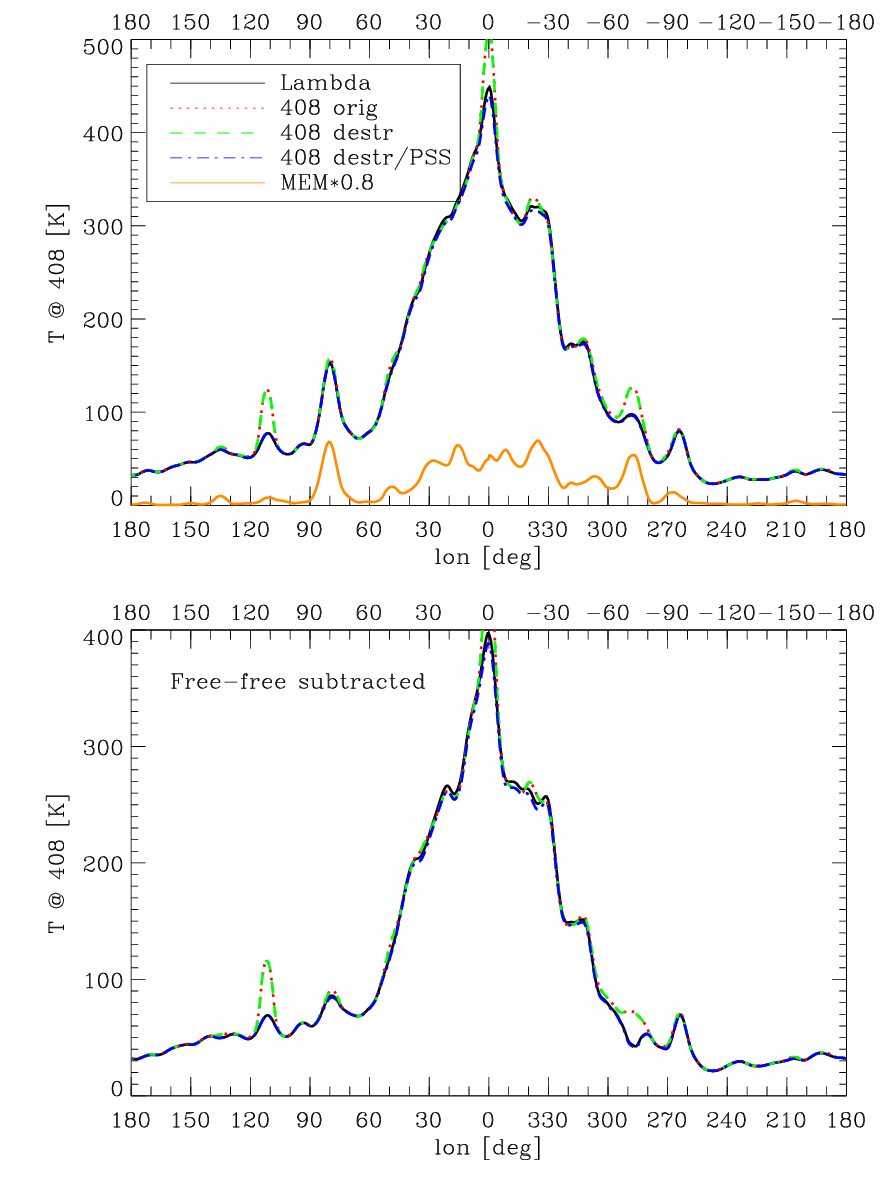}
\caption{ Impact of processing steps on 408~MHz profile.  The {\it
    top} plot shows the profiles of the LAMBDA map ({\it black}), the
  raw 408~MHz map ({\it red dotted}), the destriped map ({\it green
    dashed}), and the destriped and point-source subtracted (PSS) map
  ({\it blue dot-dashed}).  The \wmap MEM free-free profiles is
  plotted on top in {\it orange} and subtracted from each of the
  408~MHz profiles in the bottom plot.  }
\label{fig:haslam_proc}
\end{figure}

The LAMBDA version of the \citet{haslam:1982} 408~MHz map has had both
destriping and point source removal algorithms applied to the data,
while the 2.3~GHz maps has had no such processing.  The free-free
model should include significant contributions from many compact HII
regions that might have been subtracted from the Haslam map as point
sources.  To verify that this has no significant impact on our
analysis, we compare the profiles of the raw map, the destriped map,
and the destriped and point-source subtracted map in
Fig.~\ref{fig:haslam_proc}.  We also compare these locally processed
maps to the version available on the LAMBDA\footnote{{\tt
    http://lambda.gsfc.nasa.gov/}} website, which is the version used
in our analysis both for this work and in Paper I.  The profiles
clearly show where a point source was removed, but the effects of the
destriping are not visible against the strong emission on the plane.
There are only three significant point sources visible (at the
Galactic centre, at roughly $\ell \sim 290\degr=-70\degr$ and at $\ell \sim
110\degr$), two of which correspond to features in the MEM free-free
estimate.  Clearly, the point-source removal is not affecting the
free-free subtraction systematically (for example by removing many
compact HII regions which would then lead to significant
over-subtraction by MEM).  What is also clear is that leaving in point
sources in the 2.3~GHz map is not likely to have a significant effect
on the results when working as we do at large scales.

However, it is worth noting that the source removed at roughly
$290\degr=-70\degr$ lies on the Carina spiral arm tangent, and as we will show
in the results, the model appears to over-predict the synchrotron
emission there.  This mismatch is perhaps then due simply to the
over-subtraction of the free-free from the point-source removed Haslam
map.  

\subsection{Free-free correction}\label{sec:ffcor}

While the synchrotron dominates over the free-free emission over most
of the sky, it does not dominate on the Galactic plane itself at any
but the lowest of our three synchrotron frequencies.  It is because of
the free-free that we cannot simply compare the total and polarised
emission at the same frequency on the plane: if the frequency is high
enough to avoid Faraday effects, then the plane is dominated by
free-free emission.  Even at the radio frequencies we use, the
free-free makes a significant contribution and must be subtracted.

In Paper I, we used the \wmap estimate for the free-free emission
using the Maximum Entropy Method (MEM) described in \citet{gold:2009}
and references therein.  One drawback of that analysis is the fact
that it attempts to decompose the CMB sky into three components --
synchrotron, free-free, and thermal dust emission -- when there is
clear evidence that a fourth component is significant in all \wmap
bands but the highest.  This fourth component is spatially correlated
with the thermal dust emission but has a non-thermal spectrum, and it
is widely believed to be due to the electric dipole emission from
spinning dust grains.  (See, e.g., \citet{boughn:2007} or \citet{planck:2011_ame} and
references therein.)  Over the lower \wmap
frequencies, this emission has a spectrum that roughly follows a power
law with an index of $\beta\sim -2.8$ (e.g.,
\citealt{davies:2006}).  Because this index is closer to the
spectrum of synchrotron than to the $\beta=-2.15$ of free-free,
one might imagine that the anomalous component would contaminate the MEM
synchrotron solution more than the free-free solution.  But clearly,
the free-free template as well is compromised, and it is difficult to
estimate to what degree the MEM free-free solution is an
over-prediction.

The more recent \wmap seven-year analysis offers an alternative
component separation based on pixel-by-pixel MCMC analysis
\citep{gold:2010}.  This method may also explicitly fit an additional
component for the anomalous emission that can vary spectrally from
pixel to pixel, while the free-free spectrum is constrained to follow
a theoretically motivated power law \citep{dickinson:2003}.

\begin{figure}
\includegraphics[width=\linewidth]{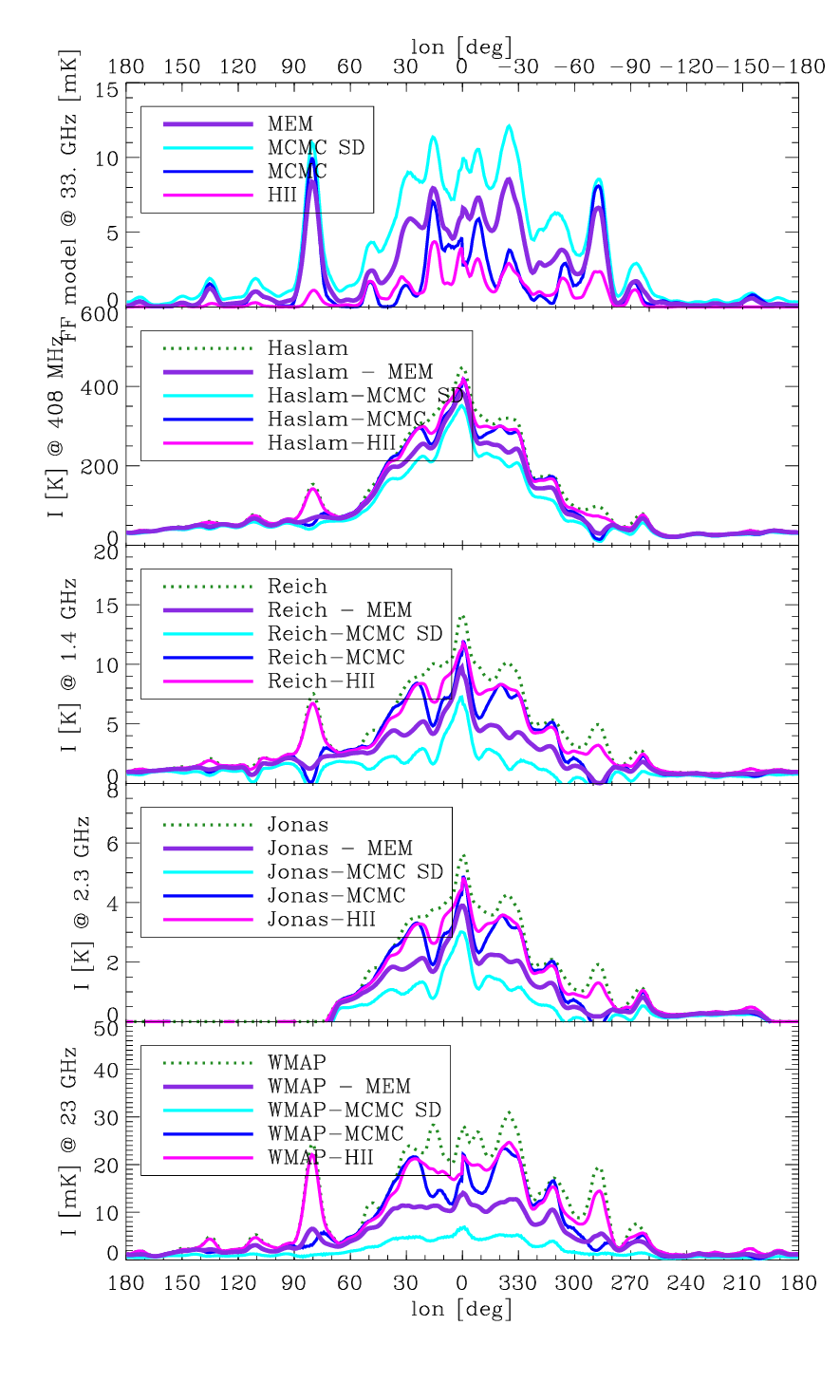}
\caption{Comparison of free-free emission models.  The \wmap team
  produce models using both the MEM and MCMC methods (both with and
  without spinning dust), while the HII model is based on the
  \citet{paladini:2003} catalog.  See text. }
\label{fig:ffcomp}
\end{figure}

\begin{figure}
\includegraphics[width=\linewidth]{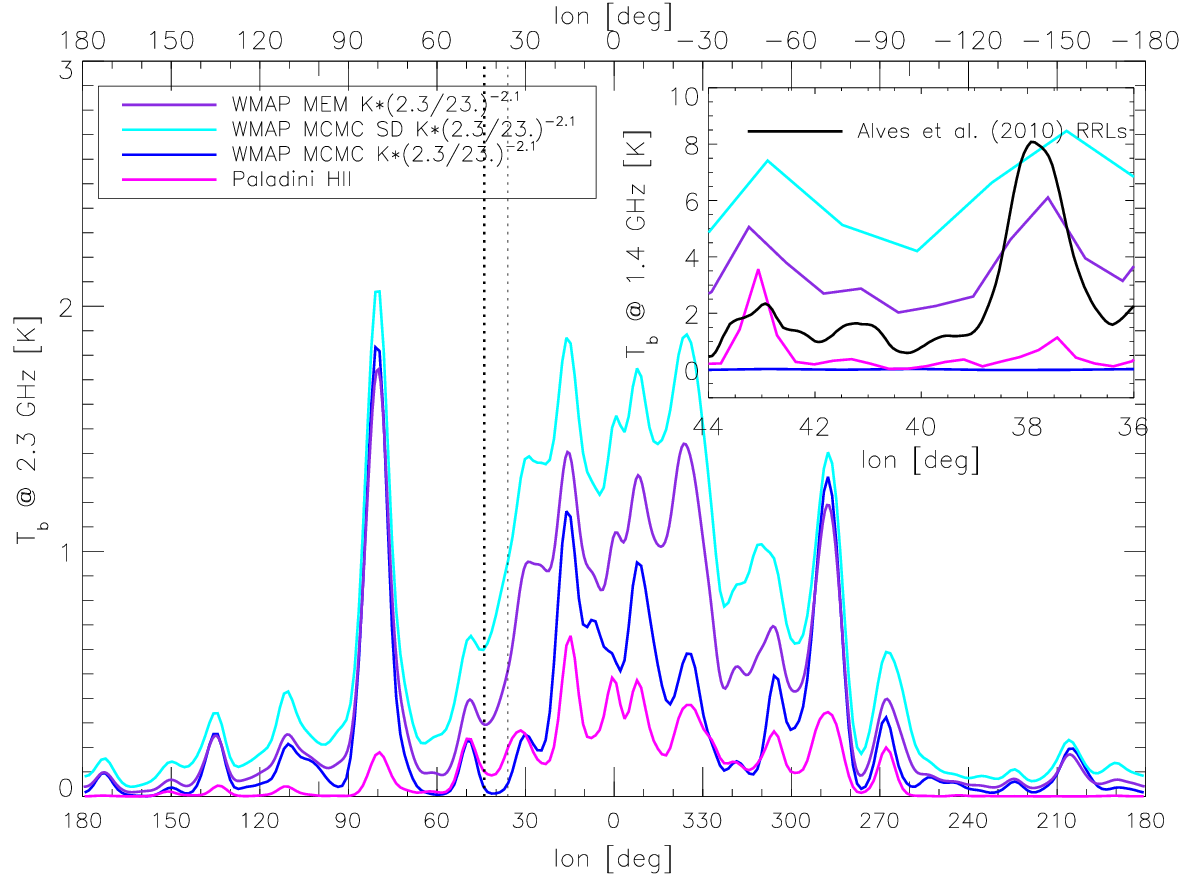}
\caption{ Comparison of the \wmap MCMC free-free predictions (with
  ({\it cyan}) and without ({\it blue}) a spinning dust component
  simultaneously fit) to the MEM solution ({\it magenta}), both
  extrapolated to 2.3~GHz, the frequency used in our analysis where
  the free-free is strongest.  Inset is a zoom, this time at 1.4~GHz,
  of the region shown in \citet{alves:2010} Fig.~10 reproduced here in
  black.  (Note that the inset shows the predictions at their full
  resolution of $1\degr$, while the main plot is smoothed to
  $6\degr$.)}
\label{fig:ff_comp}
\end{figure}

Figure \ref{fig:ffcomp} shows the total intensity profiles at each
frequency and the free-free correction for different models.  In some
regions, notably the spiral arm tangent Carina around longitude of
$290\degr=-70\degr$ and the Cygnus arm around 80$\degr$, the MCMC solutions
both clearly over-predict the low-frequency free-free and the
subtracted profile drops below zero at 1.4 and 2.3~GHz.
A vital independent estimate of the free-free emission comes from the
study of radio recombination lines (RRLs), where a comparison of
ionized hydrogen transition lines to continuum emission allows a
measurement of the free-free emission measure.  Unlike using H$\alpha$
as a tracer, this measure is not affected by dust absorption and is
the only way to measure the free-free on the plane.
\citet{alves:2010} have performed such an analysis over a small region
of the plane from 36$\degr$ to 44$\degr$ longitude.  As the most
reliable measurement of the free-free emission, it is an interesting
region in which to compare the free-free predictions based on the
microwave data.  The inset of Fig.~\ref{fig:ff_comp} shows the region
analyzed in Alves et al. compared to the models.  In that region, the
MCMC solution finds no free-free emission in the case without a
spinning dust component, while the MCMC result with spinning dust removed 
significantly over-predicts.  The MEM solution, though at lower
resolution and far from perfect, more closely reflects the emission
measured by the RRL analysis.  Alves et al. find that in the region
studied, the MEM over-predicts the free-free emission by roughly
20--30 per cent  on the plane, presumably due to contamination by the anomalous
dust emission.

We have also looked at the catalogue of HII regions by
\citet{paladini:2003} to determine the distribution of free-free
emission along the plane based on individual compact regions.  This
does not, of course, take into account the diffuse emission but
provides an interesting comparison as a clear minimum amount of
emission.  Using the catalog of positions and fluxes, Paladini et
al. created a smoothed map which we can compare to the diffuse
emission data and the other free-free models.  This is also shown in
Fig.~\ref{fig:ffcomp}.

\begin{figure}
\includegraphics[width=\linewidth]{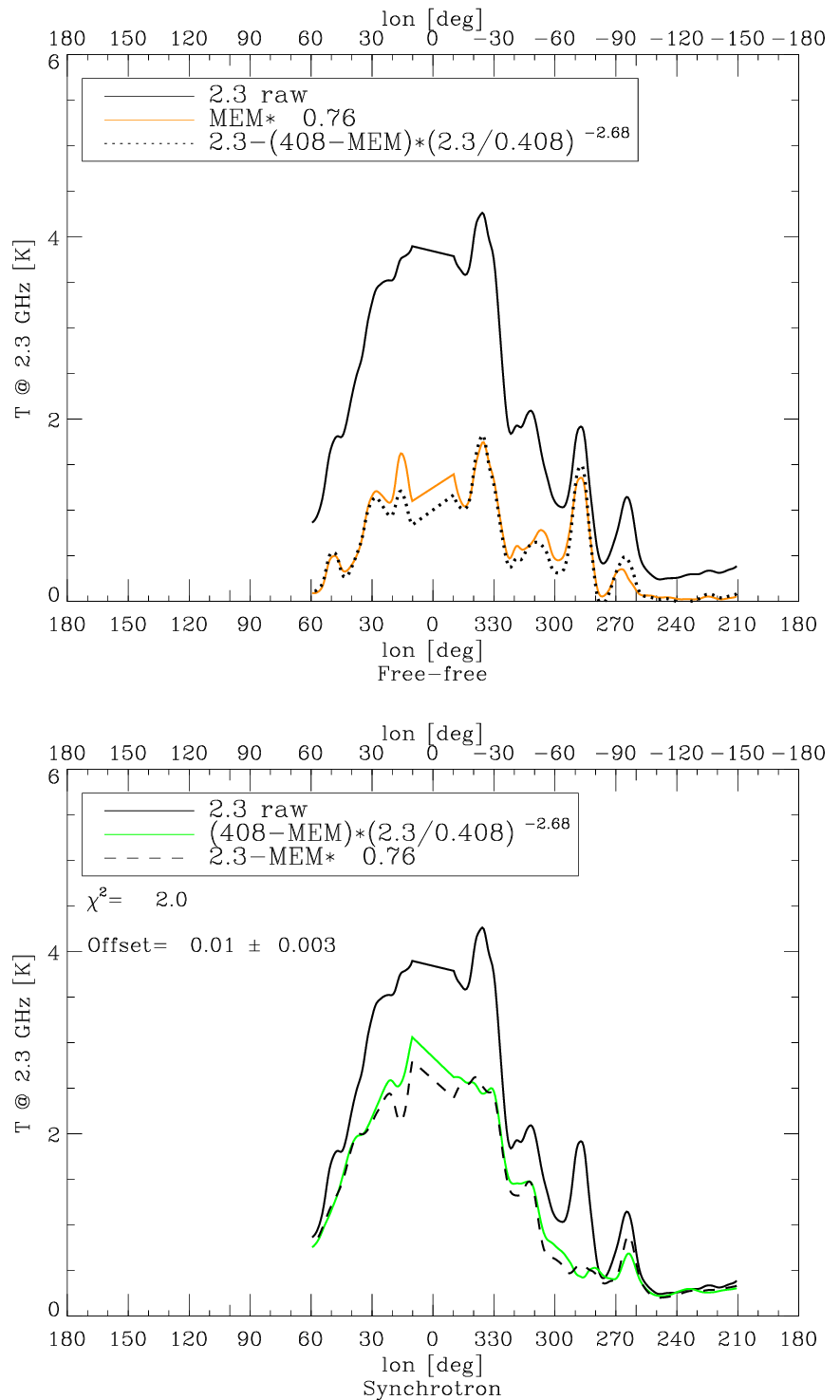}
\caption{ Comparison of 2.3~GHz survey with 408~MHz survey and the MEM
  free-free model on the plane.  In both plots, the {\it solid black}
  curve shows the full 2.3~GHz survey profile along the plane.  We
  then fit both synchrotron and free-free templates and compare the
  residuals to the scaled templates. See text.  {\it On the top} are
  shown fitted the MEM free-free solution ({\it orange}) and the
  profile after the fitted 408 synchrotron profile is subtracted ({\it
    dotted black}).  {\it On the bottom}, we see the synchrotron
  residuals ({\it dashed black}) after we subtract the fitted MEM
  compared to the fitted 408~MHz profile ({\it green}).  The scale
  factors of the fits are given in the legends, as a spectral index in
  the case of synchrotron, while the offset is printed below the
  legend.}
\label{fig:jonas_reich}
\end{figure}

We conclude that the MCMC free-free predictions appear to be less
accurate than the MEM.  In some regions, both clearly over-estimate
the free-free, while in others they clearly underestimate it.  For
this reason, we consider the MEM solution to be preferable.  As
discussed in \S~\ref{sec:offsets}, we studied the offsets of the
intermediate frequency datasets as well as the free-free correlation
by simply performing linear fits of the 408~MHz profile as a
synchrotron template and of the MEM free-free profile.  This analysis
gives not only a fitted offset but a fitted amplitude of the free-free
template compared to the reference dataset.  The coefficients are
given on the plots in Fig.~\ref{fig:jonas_reich} and show that the
free-free fits at an amplitude of lower than 80 per cent  of the MEM
prediction, consistent with what Alves et al. find.  Taking this into
account, we lower its amplitude by a conservative 20 per cent  when using it
to correct each of the total intensity profiles.  If we consider this
factor of 20 per cent  on the free-free correction to be an estimate of a
systematic uncertainty in our analysis, then it implies an uncertainty
in the spectral index of synchrotron emission from 408~MHz to 2.3~GHz
of roughly 0.05.
%


%

\section{Galaxy Modeling}\label{sec:models}

Our method follows closely that described in detail in Paper I, so we
only summarize briefly here.  We use the {\sc hammurabi} code
\citep{waelkens:2009} to simulate all the observables (synchrotron
total and polarised intensity at each frequency) based on the 3D
inputs models for the CREs and the magnetic field.  The CRE
distribution, both spatial and -- newly in this work -- spectral, is
provided by the {\sc galprop} code and \citet{strong:2010}, which
looks at Fermi $\gamma$-ray data.  Our magnetic field model is that
described in Paper I, though we again allow to vary the
$B_\mathrm{RMS}$ and $f_\mathrm{ord}$ parameters that determine the
strengths of the ordered and isotropic random components.  We compare
to the datasets described in \S~\ref{sec:obs} using a simple $\chi^2$
to estimate the likelihood of a given parameter set.  Again we use the
{\sc cosmomc} code of \citet{lewis:2002} as a generic sampler to
generate the Markov chains from which we determine the best-fitting
model parameters.  Where in Paper I we varied $B_\mathrm{RMS}$ and
$f_\mathrm{ord}$, in this work we additionally vary the spectral index
of the low-energy CRE injection spectrum.

In the following sections, we describe these changes in more detail.  

Note that since we do not vary the parameters of the coherent magnetic
field, nor the thermal electron density, in this work we do not make
further use of the rotation measure data or model prediction.  (The
polarisation frequency we use is at high enough frequency that, though
the RM is properly computed and applied in {\sc hammurabi}, it is
negligible.)  It is important, however, to remember the constraint
that these data place on the coherent field strength as demonstrated
in Paper I.  These data and model remain important to visualizing the
morphology of the field in the plane, so we continue to plot them with
our synchrotron data.

\subsection{Cosmic-ray electrons:  {\sc galprop}}\label{sec:cres}

We use the {\sc galprop} code to propagate primary electrons and
secondary positrons and electrons and thereby to model the resulting
distribution of Galactic cosmic ray electrons and positrons.  The
inputs are models of particle injection spectra (primary electrons and
protons), the smooth spatial distribution of sources, the
interstellar radiation environment, diffusion processes, and
synchrotron losses based on a model magnetic field.  The secondary
lepton source function is obtained from the CR proton
distribution.\footnote{For speed, we do not include Helium in the
  propagation, which does produce additional secondary leptons.  We
  determined that the contribution is too small to make a significant
  difference to the resulting synchrotron.}  Our baseline is the
``z04LMPDS'' model from \citet{strong:2010} with a few modifications.
We also test the ``z04LMS'' from the same paper, the main difference
being the fact that the latter includes the effects of
re-acceleration.  See \S~\ref{sec:reacc}.

We have lowered the simulation spectral range (here 0.2 to 1000 GeV)
and resolution (sampling as $E_i=E_02^i$) as much as possible to speed
up the propagation, testing with each change that the results for our
relevant observables remain unchanged, and replaced the {\sc galprop}
magnetic field model with our own.  There are several things to
  note about the magnetic field in {\sc galprop}.  Though our field
model has a trivial constant vertical structure, and though {\sc
  galprop} will use the field up to 4~kpc off the plane in the
propagation, we find that the resulting CRE distribution on the plane
is not affected by this inaccuracy.   Secondly, note that the
  small-scale B-field variations have negligible effect on the
  electron spectrum since electrons sample a very large region during
  propagation in the galaxy.  Only the strength of the total field is
  relevant for electron energy losses.  Furthermore, note that in {\sc
    galprop} there is no explicit connection between the B-field and
  the cosmic-ray diffusion coefficient, since the latter is derived
  observationally from nuclei secondary-to-primary ratios.  

  We can then explore the low-energy end of the cosmic
ray spectrum by varying the spectral index of the injected electron
spectrum below 4~GeV.  This energy range is so far ill-constrained, as
solar modulation affects local measurements of CRE densities.  (The
``z04LMPDS'' default for this parameter was $-1.6$.)

\begin{figure}
%
\includegraphics[width=\linewidth]{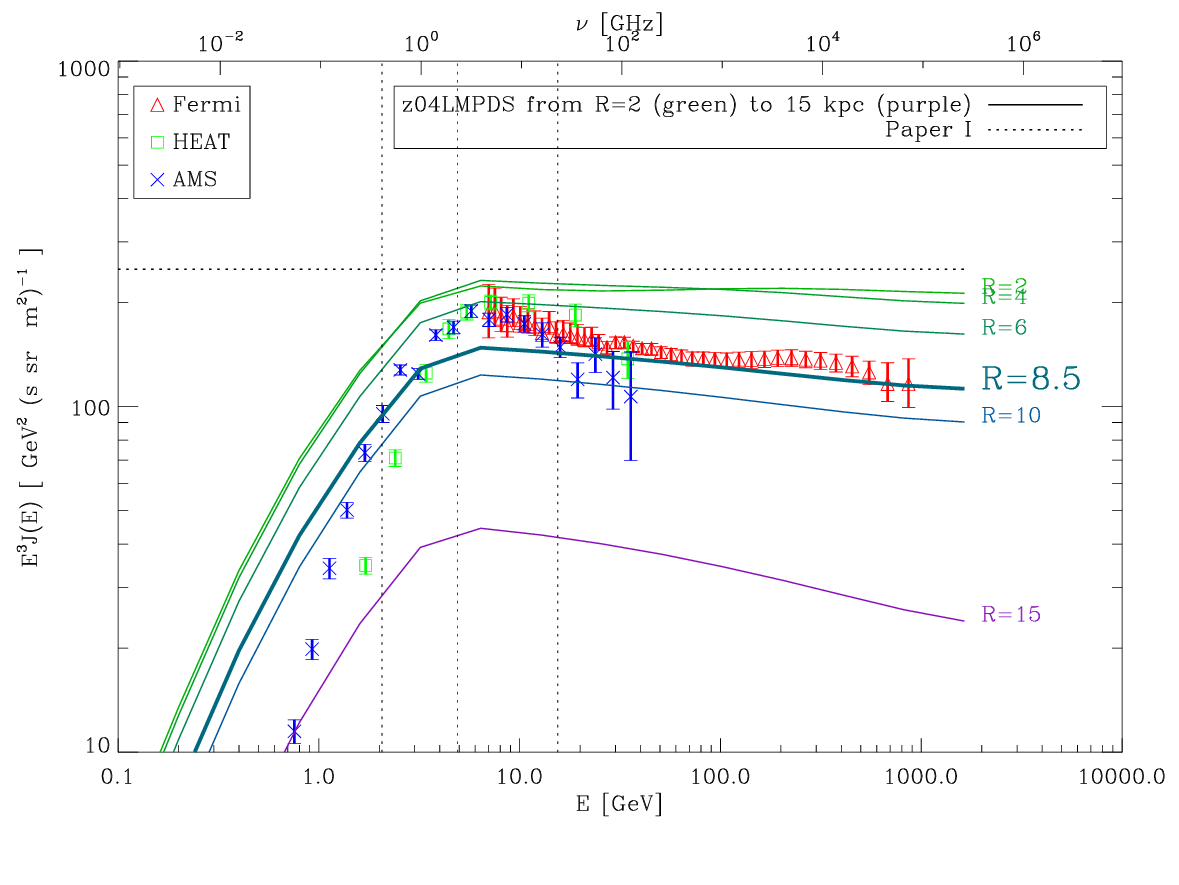}
\caption{Comparison of example {\sc galprop} cosmic ray spectra at different
  galacto-centric radii to the simple power law spectrum as used in
  Paper I.  Vertical dotted lines mark the frequencies 408~MHz,
  2.3~GHz, and 23~GHz.  Over-plotted are data from Fermi
  \citep{fermi:2010}, HEAT \citep{duvernois:2001}, and AMS
  \citep{ams:2002} using the database of \citet{strong:2009}. The top
  axis indicates approximately the effective synchrotron frequency
  corresponding to the electron energy on the lower axis, assuming a
  field strength of 6~$\mu$G; see text and Eq.~\ref{eq:webber}. Note
  that the predicted interstellar spectra are not expected to match
  the data below 10 to 20~GeV due to solar modulation; see
  \S~\ref{sec:modulation}.}
\label{fig:cres}
\end{figure}

See Table \ref{tab:galdefs} for a list of the parameters that differ
between the two models.  An example and comparison with data is shown
in Fig.~\ref{fig:cres}.

\begin{table}\begin{tabular}{lcc}
\hline
Parameter & z04LMS & z04LMPDS \\
\hline
D0\_xx & $5.8\times 10^{28}$ & $3.4\times 10^{28}$ \\
D\_g\_0 & 0.33 & 0 \\
D\_g\_1 & 0.33 & 0.5 \\
v\_Alfven & 30. & - \\
electron\_g\_1 & 2.4 & 2.3 \\
electron\_norm\_flux & $0.32\times 10^{-9}$ & $0.32\times 10^{-9}$ \\
\hline

\end{tabular}
\caption{Comparison of two {\sc galprop} models used (i.e. GALDEF contents).
  The z04LMS model includes re-acceleration, while z04LMPDS is pure diffusion.  
  Only the physical parameters are given;  other GALDEF
  differences include the spectral energy resolution, as the model with
  re-acceleration is more sensitive to the resolution at low energies, while
  without re-acceleration, the propagation can be sped up.}
\label{tab:galdefs}
\end{table}

\subsection{Integration}

In Paper I, we computed the total synchrotron emission along the line
of sight by assuming a power law distribution of CREs:
\beq \label{eq:synch} 
I_\rmn{sync}(\nu,p) \propto \int_\rmn{LOS} \rmn{d}r J_\rmn{CRE}(\mathbf{r}) B_\perp(\mathbf{r})^{\frac{p+1}{2}}
\eeq 
where $I$ is the specific intensity and $J_\rmn{CRE}(\mathbf{x})$ is
the density of cosmic ray electrons (CREs) described explicitly in
\S~\ref{sec:cres}.  The total intensity, or Stokes I, in a given
observing beam is then I$=\int Id\Omega$.  We assumed, as is commonly
done, $p=3$, where the CRE distribution is
$N(\gamma)d\gamma\propto\gamma^{-p}d\gamma$.  This then implies that $I(\nu_1)/I(\nu_2)=(\nu_1/\nu_2)^{-3}$.  

But since we are now using the full spectral information given by
{\sc galprop}, we must perform the full integration over electron energies
$\gamma$.  At each position, the total synchrotron power is:
\beq \label{eq:synch2}
P_{tot}(\mathbf{r},\omega) \propto  \int \rmn{d}x J_\rmn{CRE}(\mathbf{r},\gamma) B_\perp(\mathbf{r}) F(x)
\eeq 
where $x\equiv \frac{\omega}{\omega_c}$ and $\omega_c\equiv
\frac{3\gamma^2B_\perp}{2mc}$.  The function $F(x)$ is an integral
over modified Bessel functions.  (In our modified {\sc hammurabi}
code, we use the GNU Scientific Library\footnote{{\tt
    http://www.gnu.org/software/gsl/}} to compute these.)  See, e.g.,
\cite{rybicki:1979} for details and for the expressions for
determining the Stokes parameters; the above expression is simply to
clarify the dependencies.

\subsection{MCMC}

The likelihood exploration is performed as in Paper I with the addition 
of a third varied parameter, {\sc galprop}'s {\tt electron\_g\_0} that
defines the spectral index of the CRE {\em injection} spectrum below
4~GeV.  (This is related to the spectral index $p$ in
Eq.~\ref{eq:synch} but not the same; the injection spectrum
described by {\tt electron\_g\_0} will be subsequently modified by the
various mechanisms in the propagation.)  This does not fundamentally
change the process but simply implies that we need more samples to
obtain convergence in the likelihood distribution.  Because of the
further added computational difficulty of running a {\sc galprop}
propagation for each sample, the computation of each sample is slowed
by a factor of three.  The third parameter is not strongly correlated
with the others, so the likelihood distributions remain roughly
independent and Gaussian.

\section{Results}\label{sec:results}

\subsection{Observables}

\begin{figure*}
\includegraphics[width=\linewidth]{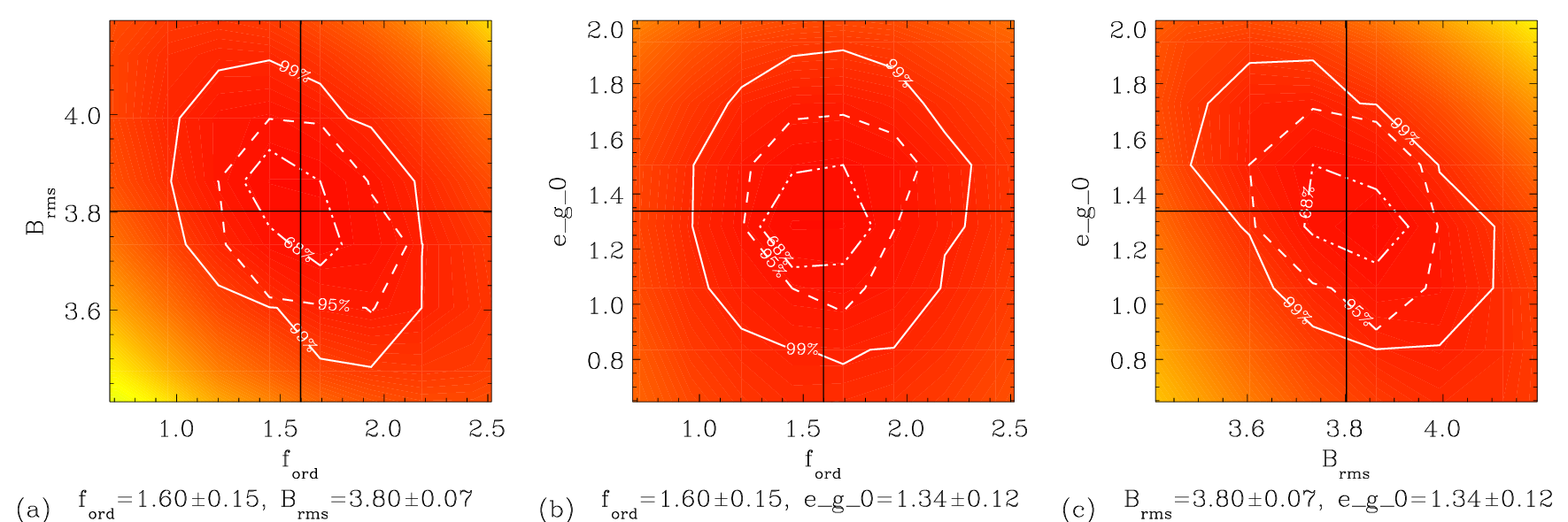}
\caption{Results of MCMC chain.  The color gives an indication of the
  sample density, while the contours give the 1, 2, and 3$\sigma$
  confidence regions.  The cross-hairs show the mean values, also
  printed below each set with their uncertainties.  For comparison,
  the Paper I results of the two-parameter fit using only 408~MHz and
  23~GHz were: $f_\mathrm{ord}=1.5\pm 0.16$ and $B_\mathrm{RMS}=2.1\pm
  0.04$.  The profiles and residuals are shown in
  Fig.~\ref{fig:residuals_bestfit}.}
\label{fig:my2dh_gp3para23}
\end{figure*}

Figure \ref{fig:my2dh_gp3para23} shows the results of fitting three
parameters to the three synchrotron profiles (total intensity at
408~MHz and 2.3~GHz, polarised intensity at 23~GHz) using the plain
diffusion z04LMPDS CR model.  Two parameters, $B_\mathrm{RMS}$ and
$f_\mathrm{ord}$, control the ratios of the three magnetic field
components (where the coherent field is fixed at the values found in
Paper I using the RM data) along with the {\tt electron\_g\_0}
parameter of {\sc galprop}, which is the spectral index of the electron
injection spectrum below the first break, i.e. for energies less than
4~GeV.

The magnetic field parameters differ from our results in Paper I due
to the fact that the CRE spectrum is different. 
Figure \ref{fig:my2dh_gp3para23} shows the new fit results, giving a
low-energy spectral index of $-1.34\pm 0.12$, which is harder than the
{\sc galprop} default of $-1.6$ usually used (e.g., \citealt{trotta:2011,fermi:2010}).

\begin{figure}
\includegraphics[width=\linewidth]{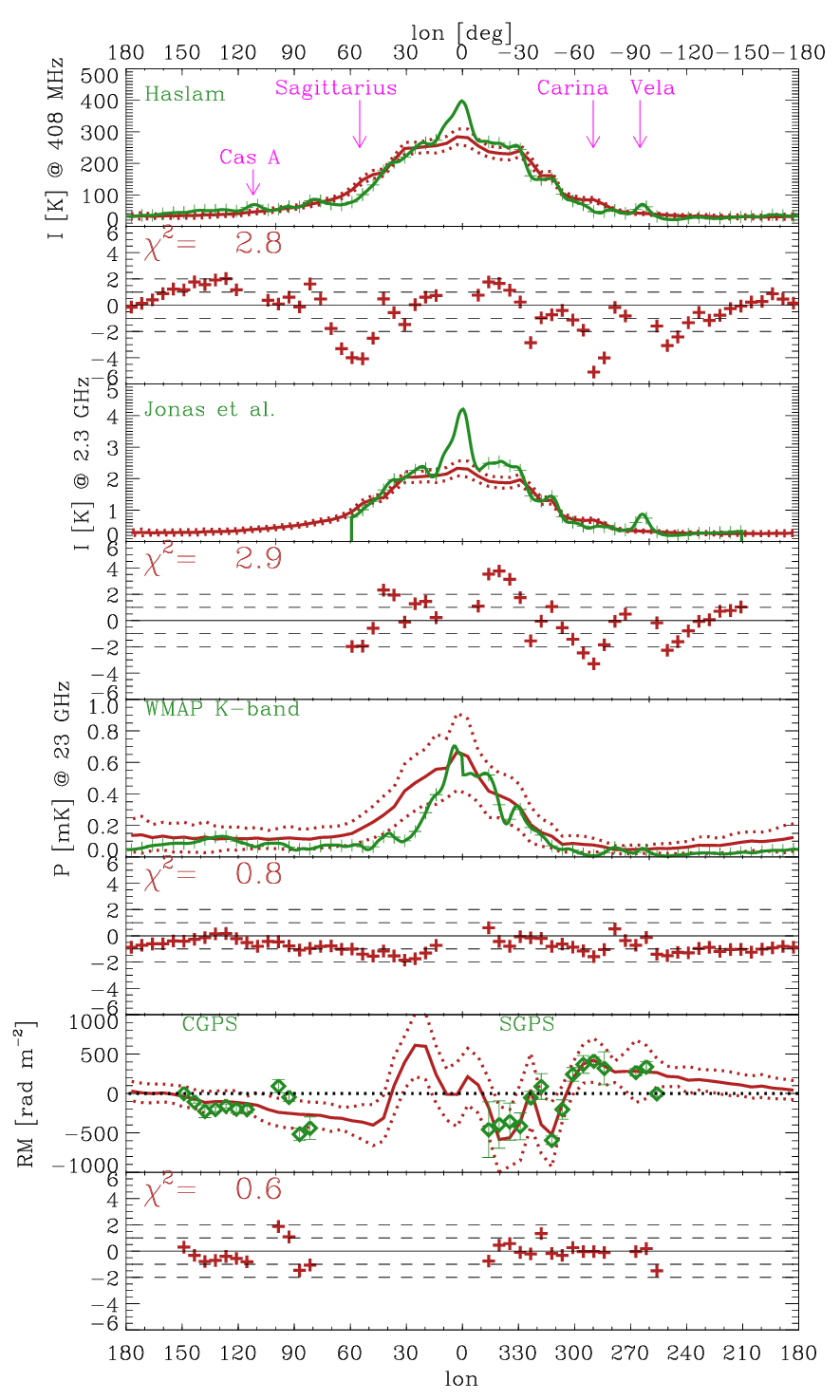}
\caption{ Best-fitting model ({\it red solid}) with its expected
  variations ({\it red dotted}) compared to the data ({\it green}) for
  each observable.  The computed $\chi^2$ excludes problem regions
  like the Galactic centre and the Cas A and Vela supernova remnants.}
\label{fig:residuals_bestfit}
\end{figure}

The fit residuals are shown in Fig.~\ref{fig:residuals_bestfit}.  The
model corresponding to the best-fitting parameters from the MCMC analysis
is shown in red compared to the data in green.  The dotted red line
represents the ``galactic variance'', i.e. the amount any individual
galaxy realization can be expected to vary from the average profile
due to its turbulent component.  In general, the profiles match well,
but there are deviations where individual arm tangents are not well
modeled (Sagittarius and Carina) or where the data contain unrelated
objects such as supernova remnants.  Though its centre is off the
plane, Vela is large enough and bright enough that it is clearly
visible in the smoothed plane profiles of both 408~MHz and 2.3~GHz.
The CasA supernova remnant is also visible around $112\degr$ longitude
in the 408~MHz total intensity profile.

As noted in \S~\ref{sec:proc}, the point source removal processing on
the Haslam map appears to have removed something in the direction of
the Carina arm tangent.  If this was a compact HII region that is
still contributing to the free-free correction, then the data used
will have been over-subtracted there.  This could explain the mismatch
between the model and data for this arm.  

It is also interesting to note that the Sagittarius arm does not fit
well, as the model significantly over-predicts the emission along that
tangent.  The \citet{cordes:2002} (a.k.a. NE2001) model for the
thermal electron distribution uses a similar spiral arm model, one of
which, the Sagittarius arm, appears to have a missing segment in
precisely the direction where our model over-predicts.  While
synchrotron is not dependent on the distribution of thermal electrons,
the rough agreement elsewhere between the synchrotron arm tangents and
the known spiral arms included in the NE2001 model does imply that the
distribution of the various components of the magneto-ionic medium are
not independent.  It does indeed appear that either the total magnetic
field or the distribution of cosmic rays also fails to follow the
modeled Sagittarius arm inward of the Sun's position.

Furthermore, as discussed above, there are new RM data by
\citet{vaneck:2011} that shed light on this issue.  Van~Eck et al. do
not attempt to fit a coherent spiral arm model similar to ours but
rather take a simpler approach to determine the field direction in
each of a set of segments.  These segments are delineated by a spiral
in the inner galaxy quadrant defined largely by negative longitudes
and by simple annuli in the rest of the inner galaxy.  With this
prescription, they can create a model to match the peaks and troughs
in all the RM data, but this model obviously has far more parameters
than our global model.  It is interesting to note, however, that the
van Eck et al. data are inconsistent with our model in the newly
covered region from roughly $20\degr$ to $60\degr$ longitude.  
  The data show a peak of roughly the same amplitude, implying the
  same coherent field strength, but shifted outward.  Clearly the
coherent field does not follow our model of the Sagittarius-Carina arm
from the first to third quadrants.  Note that the NE2001 model also
predicts a `bite' taken out of its otherwise similar spiral arm model
in the direction of Sagittarius.  To construct a model consistent with
all observations is now increasingly challenging, which is an
improvement over the previous situation of having too many possible
models.  It is also beyond the scope of this paper, though we will
address the issue in future work that will also make use of Planck
data.  Note that this inconsistency is in the geometry and not
  the strength of the magnetic field components and therefore does not
  impact the current scientific results based on the synchrotron
  spectrum.

We note that the best-fitting polarisation profile appears systematically
slightly higher than the data.  (The fact that the $\chi^2$ remains
near one is due to the large and forgiving variance from the turbulent
field component.)  It is overly simplistic to say that we can
perfectly fit our three synchrotron frequencies by varying three
parameters: $B_\mathrm{RMS}$, $f_\mathrm{ord}$, and $electron\_g\_0$.
The total magnetic field strength does have an impact on the CRE
spectrum at energies above a few GeV, where the 23~GHz synchrotron
will be affected.  To correct this bias, one might lower
$f_\mathrm{ord}$, but this would lower the total magnetic field
strength, reduce the CRE synchrotron losses, harden
the CRE spectrum above a few GeV, and thereby harden the synchrotron as
well.  Therefore we cannot correct the bias with only our three
parameters.  The implication is that the {\sc galprop} parameter {\tt
  electron\_g\_1} determining the injection spectral index between
4~GeV and 1~TeV should be steeper than the $-2.25$ found in
\citet{strong:2010} with a different magnetic field model.  If we had
unlimited computational resources, we should simultaneously explore
more of the {\sc galprop} parameters and include Fermi CRE and $\gamma$-ray
data, but this is not currently feasible.

%

\subsection{Synchrotron spectral index}

\begin{figure}
\includegraphics[width=\linewidth]{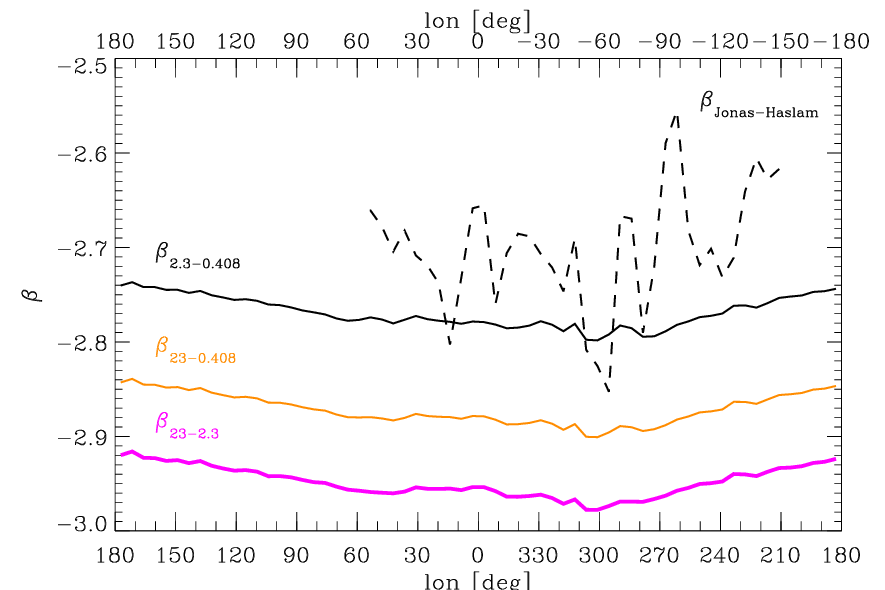}
\caption{ The best-fitting model's synchrotron spectral indices between each of the
  set of synchrotron frequencies.  For comparison, the spectral index
  from Haslam 408~MHz to Jonas et al. 2.3~GHz is over-plotted.  See
  text.}
\label{fig:betas}
\end{figure}

Fig.~\ref{fig:betas} shows the synchrotron spectral index between each
pair of frequencies as a function of longitude.  The implied
synchrotron spectral index from 408~MHz and 2.3~GHz is roughly
$-2.8<\beta < -2.74$, and that between 2.3~GHz and 23.~GHz is $-2.91 <
\beta < -2.98$ depending on the longitude.  Previous synchrotron
spectral index analyses that don't explicitly exclude the plane
\citep{reich:1988,davies:1996, giardino:2002,platania:2003} tend not
to discuss it in detail either because of the uncertainties in the
free-free contamination or simply because they are primarily
interested in high latitudes.  Most do conclude with a low-frequency
spectral index of roughly $-2.8$ and a high frequency index of roughly
$-3$.


Recall that our method computes the synchrotron emissivity at each
frequency and at each Galactic position independently based on the
input spatial and spectral distribution of CRE injection sources, the
interstellar radiation field (ISRF), and of course the Galactic
magnetic field.  Because all of the other components are azimuthally
symmetric about the Galactic centre, the structure seen in
Fig.~\ref{fig:betas} is a result of the changing magnetic field
strength when looking in different directions.  This affects the
synchrotron spectral index as can be seen both via
Eq.~\ref{eq:synch2} as well as more simply via an approximation of
the effective synchrotron frequency for an electron of a given energy:
\beq\label{eq:webber}
\nu_\mathrm{eff}\text{\footnotesize{[MHz]}}=16 B_\perp \text{\footnotesize{[$\mu$G]}} E^2 \text{\footnotesize{[GeV$^2$]}}
\eeq
from \citet{webber:1980}.  The steep spectral break at 4~GeV means
that a change in the field strength changes the region of the electron
index that the radio synchrotron frequencies are effectively sampling,
which in turn can have a large effect on the predicted synchrotron
spectral index.

The structure of each of the synchrotron index profiles shows the
hardest emission toward the anti-centre and the softest emission toward
$\ell\sim 300\degr=-60\degr$.  The Galactic centre region $-60\degr
\lesssim \ell\lesssim 50\degr$ is bracketed asymmetrically by
softer-spectrum dips on either side, which hints at the source being
the spiral magnetic field morphology.  Such an asymmetry is expected
for a modulation related to the strong Scutum-Crux and Perseus arms.
The latter, passing behind the Solar position would explain the
anti-centre hardening.  (Recall from Paper I that the results
indicated that the Sagittarius-Carina arm is very weak in terms of the
magnetic field strength, and the Norma arm not well constrained.)
Note that a similar figure produced using a simple exponential disc
magnetic field model, e.g. the {\sc galprop} default, shows no such
structure; instead there is only a very slight hardening toward the
Galactic centre as one would expect.
%

The measured spectral index between the two total intensity
frequencies is also shown in Fig.~\ref{fig:betas} as the black dashed
line.  Its variation along the plane (due to the turbulent variations
in the magnetic field) is larger than the predicted {\em average}
features of Fig.~\ref{fig:betas}, so it is unlikely we can ever
confirm such structures.  We can also see that there is again a small
bias in that spectral index; the best-fitting model returns an average
spectral index between 0.408 and 2.3~GHz of roughly $-2.75$, while the
data show a slightly harder average of around $-2.7$.  As discussed
above, this is due to the fact that the three parameters varied do not
have complete freedom to fit the data perfectly due to the complicated
interaction of the magnetic field strength and the cosmic ray
distribution in determining the resulting synchrotron.  If we had the
computational resources to vary more of the {\sc galprop} parameters (and
necessarily include the Fermi electron and $\gamma$-ray data which
also constrain them), we would expect a model with a slightly harder
low-frequency spectrum and a slightly steeper high-frequency spectrum.
(This can just be seen by eye in Fig.~\ref{fig:residuals_bestfit} as a
slight positive bias in the model at 408~MHz.)  Though we have varied
the low-energy electron spectrum, we have not varied the high-energy
end, which is the source of the bias.

\subsection{CRE spectrum}

\begin{figure}
\includegraphics[width=\linewidth]{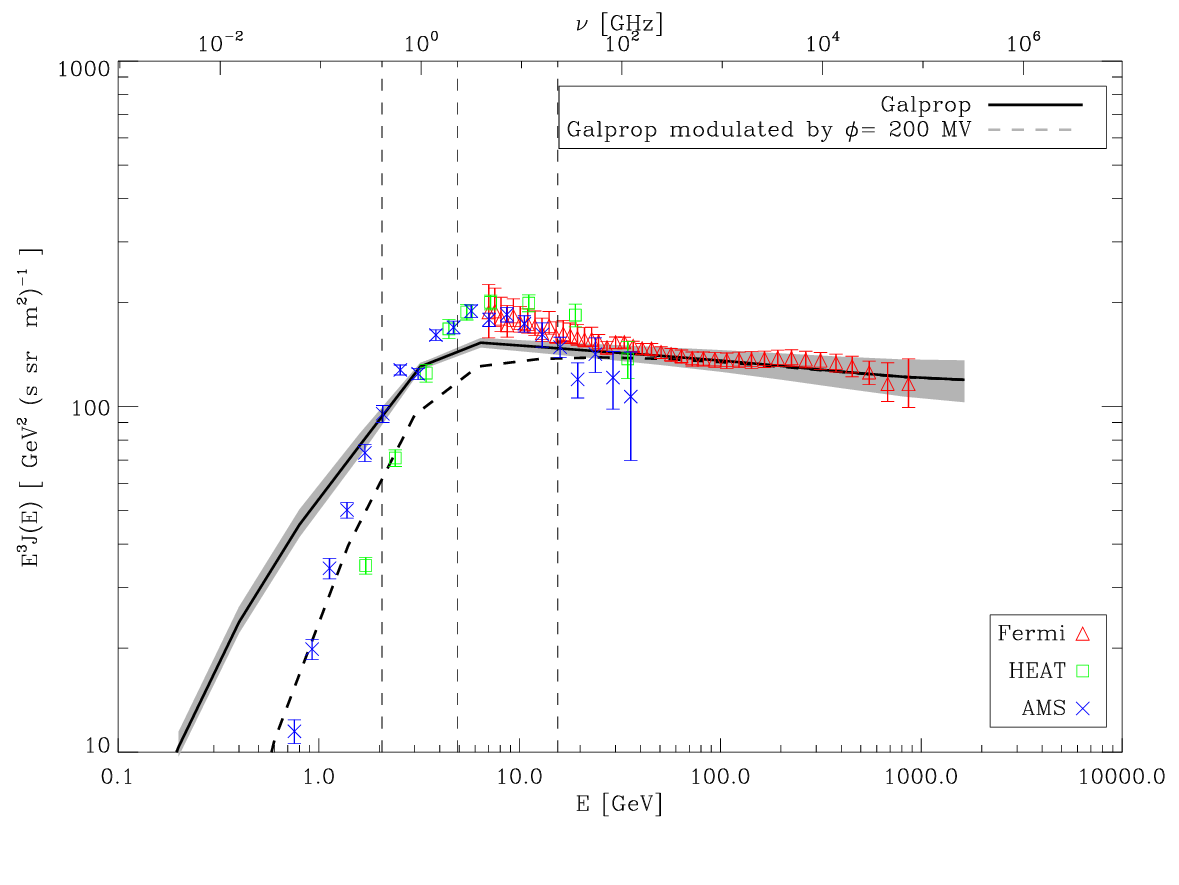}
\caption{ Best-fitting CRE spectral prediction for the Solar
  neighborhood ({\it solid black}) compared to Fermi, HEAT, and AMS
  data.  The grey shading indicates the region of uncertainty: at low
  energies, bounded by the 1$\sigma$ uncertainty in the electron
  injection spectral index, and at high energies, due to the variance
  in the local magnetic field.  The {\it dashed black} line shows an
  attempt to modulate this spectrum with the \citet{gleeson:1968}
  prescription using $\phi=200$~MV, which clearly does not work at
  all energies.  Lastly, note that this spectrum is {\em not} a fit to the
  Fermi CRE data but rather uses parameters constrained with Fermi
  $\gamma$-ray data and yet shows good consistency at high energies.}
\label{fig:cre_results}
\end{figure}

The resulting CRE spectrum is shown in Fig.~\ref{fig:cre_results}.  At
higher energies, above roughly 20~GeV, it is consistent with the Fermi
data.  Below these energies, the local measurements are not directly
comparable to the model due to solar modulation.  The best-fitting value
of the low-energy electron injection spectrum gives $J(E)\propto
E^{-1.34}$, which is slightly harder than the previous {\sc galprop} default
value of $-1.6$.  That figure also shows in shaded grey the uncertainty
in the low-energy end of the spectrum due to the error bar on that
index as estimated by our MCMC chains.  The fact that this uncertainty
appears so small is due to the sensitive dependence of the synchrotron
amplitudes and our fairly precise fitting on the plane.  The shaded
grey at high energies shows the uncertainty due to a different source.
The parameters are fit by comparing a synchrotron profile along the
plane, which averages through the galaxy, to many independent
simulations of a magnetic field that includes a stochastic component.
What we are plotting in this figure is the predicted cosmic ray
spectrum at the solar position compared to locally measured CREs.  But
that local measurement is affected by any fluctuation in the local
magnetic field.  To account for this uncertainty, we compute a mean
spectrum at the solar radius and the corresponding variance to give an
idea of how much the high-energy end of the local spectrum can vary due to
localized magnetic field fluctuations.  (This affects only the higher
energy end of the spectrum, because it is at high energies that
synchrotron energy losses are most severe and therefore most affected
by local fluctuations.)


The uncertainty in the CRE injection spectral index returned by the
MCMC analysis is the statistical uncertainty due to the ``galactic
variance'' from the turbulent field component.  There remains a
systematic uncertainty due to the free-free correction discussed in
\S~\ref{sec:ffcor}.  This effect, however, is smaller than the
galactic variance, as it implies a possible error on the synchrotron
index of only 0.05.

\subsection{Re-acceleration}\label{sec:reacc}

The ``z04LMS'' model from \citet{strong:2010} includes diffusive
re-acceleration.  The addition of this process to the propagation
helps to reproduce an observed peak in the secondary to primary
particle ratio, though the evidence for the process as implemented in
z04LMS is not unambiguous; see \citealt{strong:2007}.  The effect of
this process on the total interstellar CRE spectrum is an additional
bump around a few GeV.  This region of the spectrum is responsible for
synchrotron bands at 408~MHz and 2.3~GHz.

Using the parameters of the ``z04LMS'' model and varying only the
spectral index of the injection spectrum below 4~GeV, we are unable to
find a fit consistent with the synchrotron data.  The MCMC tend toward
{\tt electron\_g\_0} of zero, which is simply because that is the only
way to counter the effect of the re-acceleration bump and leave a CRE
spectrum in the few GeV range that is consistent with the radio
synchrotron data.  
Either the re-acceleration is not correct, or we must vary other
parameters such as those defining the diffusion, which would then risk
a result inconsistent with the $\gamma$-ray and other data that model
was made to fit.

\subsection{Solar modulation}\label{sec:modulation}

Though various satellites have directly measured the flux of cosmic
ray electrons near the Earth, these measurements do not reflect the
average distribution of particles in the ISM.  The reason is the
interaction of the CREs with the solar wind.  In essence, there are
magnetic irregularities carried with the solar wind that scatter lower
energy CREs.  Not only is it difficult to quantify the modulation of
the CRE spectrum, but the modulation is time-varying.
\citet{gleeson:1968} give a simple prescription based on a single
parameter for modulating a CRE spectrum using the force field
approximation to account for this solar system effect.  Though a
simple approximation, this prescription remains in use
\citep{fermi:2010,trotta:2011} for lack of a significantly better
alternative.  Using this method, \cite{duvernois:2001} estimated 755
MV for 1994 and 670 MV for 1995, the relevant time periods for the
HEAT observations, while \cite{ams:2002} estimate $650\pm 40$~MV for
their 1998 observations.  These figures, however, generally depend on
an assumed interstellar CRE spectrum with a steeper low-energy CRE
injection spectral index ($-2.1$), which is now ruled out by the
synchrotron data.

The only way to study the CRE spectrum at lower energies, then, is to
use a tracer such as synchrotron emission or inverse-Compton
$\gamma$-rays. Unfortunately, on the Galactic plane, the $\gamma$-ray
spectrum is dominated by the gas component rather than the inverse
Compton component (and that is in turn dominated by higher energy CREs
or nuclei) so that the $\gamma$-ray data cannot help us much with the
low-energy CRE spectrum.  It is certainly possible to produce plots of
modulated theoretical CRE spectra matching even the low-E data such as
Fig.~22 of \cite{fermi:2010}.  However, as discussed in
\S~\ref{sec:reacc}, this is due to a combination of adding an
uncertain re-acceleration process and using a simplistic solar
modulation prescription.  It is unclear that we are learning anything
meaningful about either process.

Examining the synchrotron emission is therefore a more reliable way to
explore the low-energy CRE spectrum.  With enough frequencies, we can
overcome degeneracies with the magnetic field parameters and obtain a
more accurate determination of the CRE spectral index below a few GeV.
It would be easier if we could use even lower-frequency synchrotron
data, but on the plane, emission much lower than 408~MHz is going to
be absorbed by ionized gas. 

It is therefore interesting to note that in
Fig.~\ref{fig:cre_results}, we see that for our best fit parameters,
the simple Gleeson \& Axford solar modulation prescription does not
match the local observations at low energies.  For lower values of
$\Phi$, the spectrum at the lowest energies would be too high, while
at the value of 200~MV as shown, the predicted spectrum is already
too low at intermediate energies of roughly 3 to 20\,GeV.  To match
all of the data without re-acceleration would therefore require a
different solar modulation model.  Likely this is an unsurprising
indication that a more sophisticated treatment of the modulation is
needed or that the cosmic ray propagation needs to be modified in the
few GeV regime.

The spectrum in Fig.~\ref{fig:cre_results} also shows that the local
measurements are in excess of the the interstellar prediction around
the region of the break at 4~GeV.  Even a more accurate solar
modulation model is unlikely to fix this problem.  
%

\section{Conclusions}

We have used a self-consistent method of modeling both the Galactic
magnetic field and the cosmic ray electron spectrum in the Galactic
plane and used comparisons with synchrotron data to constrain physical
parameters of the model.  The particular advantage of this analysis
over previous analyses is our ability to constrain simultaneously the
CREs and magnetic field in the plane while taking into account the
field's effect on the CRE propagation.  Using multiple synchrotron
frequencies and the combination of total intensity and polarised
intensity, we can constrain not only the magnetic field parameters as
in Paper I but also begin to provide among the first constraints on the
low-energy CRE spectrum.  This regime of the CRE distribution is not
accessible via either inverse-Compton $\gamma$-ray production or
through direct measurement, since the locally measured distribution is
affected by solar modulation.

We find that the magnetic field parameters measured in Paper I must be
modified slightly when a realistic CRE spectrum is taken into account.
In particular, the random component is even larger compared to the
coherent component, with $B_\mathrm{RMS}=3.8$ compared to 2.1 in our
previous work (see \citealt{jaffe:2010} for the field
parametrization).  The large change illustrates both the significance of
dropping the assumption of a single CRE spectral index from below a
GeV to $10^2$ GeV, which was not realistic, and the fact that our
fitting returns a much harder low-energy spectral index requiring a
large magnetic field change to compensate.  This does not change the
main point of that first paper, which was to demonstrate the
importance of the ordered component of the magnetic field, since the
ratio of the ordered component to the isotropic random component
remains roughly unchanged.  The implication is an even stronger
turbulent Galactic magnetic field; the isotropic component has a peak
RMS of nearly 8~$\mu$G along the arm ridges, while the total field
strength on the ridge is then as much as 15~$\mu$G compared to 
roughly 5~$\mu$G in the inter-arm regions.

The fitting of the low-energy cosmic ray injection spectrum gives us a
slightly harder spectrum of than is generally assumed.  The {\sc galprop}
parameter {\tt electron\_g\_0} representing the spectral index of the
injection spectrum below 4~GeV, is determined to be $1.34\pm 0.12$,
though the oft-used value of 1.6 is not strongly ruled out.  Our more
robust measure of the index implies a somewhat harder injection
spectrum and confirms the strong break at a few GeV between the low
energy regime with an index of $-1.3$ and the higher energy range, with
an index of roughly $-2.3$.  We find the re-acceleration model to be
incompatible with the low-frequency synchrotron data for any value of
the CRE spectral index.  Lastly, we find that the shape of the
predicted CRE spectrum combined with the force field approximation to
the solar modulation is not able to reproduce the local measurements
of the CRE density.  More work is needed to find a model consistent
with all the available datasets.

\section*{Acknowledgments}

We gratefully acknowledge useful discussions with C. Evoli,
C. Dickinson, R. Davies, and W. Reich.  We thank J. Jonas for the
2326~MHz data as well as M. Alves for the RRL data plotted in
Fig.~\ref{fig:ff_comp} and R. Paladini for the HII data.
We acknowledge use of the HEALPix software \citep{healpix} for some of
the results in this work.  We acknowledge the use of the Legacy
Archive for Microwave Background Data Analysis (LAMBDA). Support for
LAMBDA is provided by the NASA Office of Space Science.
This research was supported by the Agence Nationale de la Recherche (ANR-08-CEXC-0002-01).

\bibliographystyle{mn2e}
\bibliography{references}



\bsp

\label{lastpage}

\end{document}